\documentstyle[11pt,twoside,epsfig,rotate]{article}

\newlength{\dinwidth}
\newlength{\dinmargin}
\begin{document}
\newcommand {\gapprox}
   {\raisebox{-0.7ex}{$\stackrel {\textstyle>}{\sim}$}}
\newcommand {\lapprox}
   {\raisebox{-0.7ex}{$\stackrel {\textstyle<}{\sim}$}}
\newcommand{\n}{1.19\pm 0.06 (stat.) \pm0.07 (syst.)}
\newcommand{\ALPHA}{1.10\pm0.03 (stat.) \pm0.04 (syst.)}
\newcommand{\nel}{1.30\pm 0.08 (stat.) \pm0.16 (syst.)}
\newcommand{\xpomlo}{3\times10^{-4}}
\newcommand{\xpomup}{0.05}
\begin{titlepage}
\begin{flushleft}
{\tt DESY 95-36}\hfill {\tt ISSN 0418-9833} \\
{\tt February 1995}
\end{flushleft}
\vspace*{4.0cm}
\begin{center}\begin{LARGE}
{\bf
First Measurement of the  \\
Deep--Inelastic Structure   \\
of Proton Diffraction}\\
\vspace*{2.5cm}
H1 Collaboration \\
\vspace*{2.5cm}
\end{LARGE}
{\bf Abstract}
\begin{quotation}
\noindent
A measurement is presented, using data taken with the H1 detector at
HERA, of the contribution of diffractive interactions to
deep--inelastic electron--proton ($ep$)
scattering in the kinematic range
$8.5<Q^2<50\,{\rm GeV}^2$, $2.4\times10^{-4}<$~Bjorken--$x<0.0133$,
and $3.7\times 10^{-4}<x_{I\!\!P}<0.043$. The diffractive
contribution to
the proton structure function $F_2(x,Q^2)$ is evaluated as a function of
the appropriate deep--inelastic scattering variables $x_{I\!\!P}$, $Q^2$, and
$\beta$ ($=x/x_{I\!\!P}$)
using a class of deep--inelastic $ep$ scattering events with no hadronic
energy flow in an interval of pseudo--rapidity adjacent to the proton beam
direction. The dependence of this contribution on
$x_{I\!\!P}$ is measured to be $x_{I\!\!P}^{-n}$ with $n=\n$ independent
of $\beta$ and $Q^2$, which is consistent with both a diffractive
interpretation and
a factorisable $ep$ diffractive cross section.
A first measurement of the deep--inelastic structure of the pomeron
in the form of the $Q^2$ and $\beta$ dependences of a factorised structure
function is presented.
For all measured $\beta$, this structure function is observed to be
consistent with scale invariance.

\end{quotation}
\vspace*{2.0cm}
{\it Submitted to Physics Letters $\mbox{\boldmath $B$}$}  \\
\vfill
\cleardoublepage
\end{center}
\end{titlepage}
\begin{Large} \begin{center} H1 Collaboration \end{center} \end{Large}
\begin{flushleft}
 T.~Ahmed$^{3}$,                  
 S.~Aid$^{13}$,                   
 V.~Andreev$^{24}$,               
 B.~Andrieu$^{28}$,               
 R.-D.~Appuhn$^{11}$,             
 M.~Arpagaus$^{36}$,              
 A.~Babaev$^{26}$,                
 J.~Baehr$^{35}$,                 
 J.~B\'an$^{17}$,                 
 Y.~Ban$^{27}$
% ...
\end{flushleft}
\begin{flushleft} {\it
% ...
$\:^3$ School of Physics and Space Research, University of Birmingham,
                             Birmingham, UK$^ b$\\
% ...
$\:^7$ Physics Department and IIRPA,
         University of California, Davis, California, USA$^ e$ \\
% ...
$ ^{11}$ DESY, Hamburg, Germany$^a$ \\
% ...
$ ^{13}$ II. Institut f\"ur Experimentalphysik, Universit\"at Hamburg,
                                     Hamburg, Germany$^ a$  \\
% ...
$ ^{17}$ Institute of Experimental Physics, Slovak Academy of
                Sciences, Ko\v{s}ice, Slovak Republic$^ f$\\
% ...
$ ^{24}$ Lebedev Physical Institute, Moscow, Russia$^ f$ \\
% ...
$ ^{26}$ Max-Planck-Institut f\"ur Physik,
                                            M\"unchen, Germany$^ a$\\
 $ ^{27}$ LAL, Universit\'{e} de Paris-Sud, IN2P3-CNRS,
                            Orsay, France\\
 $ ^{28}$ LPNHE, Ecole Polytechnique, IN2P3-CNRS,
                             Palaiseau, France \\
% ...
$ ^{35}$ DESY, Institut f\"ur Hochenergiephysik,
                              Zeuthen, Germany$^ a$\\
 $ ^{36}$ Institut f\"ur Teilchenphysik,
          ETH, Z\"urich, Switzerland$^ i$\\
% ...
$ ^a$ Supported by the Bundesministerium f\"ur
                                  Forschung und Technologie, FRG
 under contract numbers 6AC17P, 6AC47P, 6DO57I, 6HH17P, 6HH27I, 6HD17I,
 6HD27I, 6KI17P, 6MP17I, and 6WT87P \\
 $ ^b$ Supported by the UK Particle Physics and Astronomy Research
 Council, and formerly by the UK Science and Engineering Research
 Council \\
% ...
$ ^e$ Supported in part by USDOE grant DE F603 91ER40674\\
 $ ^f$ Supported by the Deutsche Forschungsgemeinschaft\\
% ...
$ ^i$ Supported by the Swiss National Science Foundation\\

   } \end{flushleft}
%

\section{Introduction and Formalism}
\label{sec-intro}

Recent measurements at HERA of deep--inelastic electron--proton ($ep$)
scattering (DIS) in the low Bjorken--$x$ kinematic range
$5<Q^2<120\,{\rm GeV}^2$ and $10^{-4}<x<10^{-2}$
have demonstrated the existence of a distinct class of events
in which there is no hadronic energy flow in an interval of (laboratory frame)
pseudo--rapidity $\eta$ adjacent to the proton beam
direction~\cite{Zrapgap,H1rapgap}. Our present understanding of DIS could be
inadequate at low $x$ because additions to the leading order QCD--based
partonic
picture are likely to be substantial. A natural interpretation of these so
called ``rapidity gap" events is based on the hypothesis that the
deep--inelastic scattering process involves the interaction of the
virtual boson probe with a colourless component of the proton. Hence there is
no
chromodynamic radiation in the final state immediately adjacent to the
direction
of the scattered proton or any proton remnant.
Observed distributions of such events are found to be consistent
with simulations based on models in which the virtual boson--proton interaction
is  diffractive~\cite{H1rapgap}, that is in which the colourless component of
the proton is hypothesised to be a pomeron ($I\!\!P$) and the virtual
boson--proton interaction may be understood as $I\!\!P$ exchange.
The observation of these rapidity gap events in DIS means that a measurement of
any short distance sub--structure of this colourless component of the proton is
possible, and thus, if the process is diffractive, of the $I\!\!P$. An
understanding of the sub--structure of this colourless component,
whether in the form of a partonic interpretation~\cite{Low,Nussinov,IngSchl}
or otherwise, is essential for further understanding of the partonic picture of
deep--inelastic lepton--nucleon scattering.

In this paper we present a measurement which quantifies the contribution of
rapidity gap events to the inclusive deep--inelastic structure function
$F_2$ of the proton. The results here follow our first measurement of this
contribution in terms of a ``diffractive structure function" $F_2^D$ as a
function of the two DIS variables $x$ and $Q^2$ reported in~\cite{H1F293}. The
measurement presented here is made as a function of the three kinematic
variables
$\beta$, $Q^2$, and $x$, or equivalently $\beta$, $Q^2$, and
$x_{I\!\!P}$, which are defined as follows:
\begin{equation}
x=\frac {-q^2}{2P\cdot q}\,\,\,\,\,\,\,\,
\,\,\,\,\,\,\,\,\,\,\,\,\,\,\,\,\,\,
x_{I\!\!P} = \frac{q\cdot (P-P')}{q\cdot P}\,\,\,\,\,\,\,\,\,
\,\,\,\,\,\,\,\,\,\,\,\,\,\,\,\,\,\,
Q^2 = -q^2\,\,\,\,\,\,\,\,\,
\,\,\,\,\,\,\,\,\,\,\,\,\,\,\,\,\,\,
\beta = \frac{-q^2}{2q\cdot (P-P')} .
           \label{eq:definition}
\end{equation}
Here $q$, $P$ and $P'$ are the $4$--momenta of the virtual boson, the incident
proton, and the final state colourless remnant respectively. The latter can be
either a nucleon or higher
mass baryonic excitation, and, if the proton interaction is
diffractive, it must be a proton or a proton excitation with isospin
$(I,I_3)=(\frac{1}{2},+\frac{1}{2})$.
Note that
\begin{equation}
x=\beta x_{I\!\!P}.
           \label{eq:trivial}
\end{equation}
It is convenient to write $x_{I\!\!P}$ and $\beta$ above as
\begin{equation}
x_{I\!\!P}  = \frac{Q^2+M_X^2-t}{Q^2+W^2-M_p^2}
        \approx \frac{Q^2+M_X^2}{Q^2}\cdot x =x_{I\!\!P/p}
           \label{eq:xpomQ2}
\end{equation}
\begin{equation}
\beta = \frac{Q^2}{Q^2+M_X^2-t} \approx \frac{Q^2}{Q^2+M_X^2}
=x_{q/I\!\!P}
           \label{eq:beta}
\end{equation}
where $M_X$ is the invariant mass of the hadronic system excluding the
colourless remnant, $t=(P-P')^2$ is the
$4$--momentum transfer squared at the incident proton vertex, and $W$ is the
total
hadronic invariant mass. In the kinematic domain of these measurements
($M_p^2\ll Q^2$, $M_p^2\ll W^2$)
and if $\mid\!t\!\mid$ is small
($\mid\!t\!\mid\ll Q^2$, $\mid\!t\!\mid\ll M_X^2$) approximating to ``the
proton's infinite momentum
frame",
$x_{I\!\!P}$ may be interpreted as the fraction $x_{I\!\!P/p}$ of the
$4$--momentum of the proton carried by the interacting $I\!\!P$ (or meson for
non--diffractive contributions), and $\beta$ as the fraction $x_{q/I\!\!P}$ of
the $4$--momenta of the $I\!\!P$ (or meson) carried by the quark interacting
with the virtual boson.

The diffractive structure function $F_2^{D(3)}$, which is a function of three
kinematic variables, is derived from a structure function $F_2^{D(4)}$ which is
a function of the four kinematic variables $x$, $Q^2$, $x_{I\!\!P}$ and $t$.
$F_2^{D(4)}$ is defined by analogy with the decomposition of the
unpolarised total $ep$ cross section. In the $Q^2$ range of these measurements,
the cross section for the process $ep\rightarrow eXp$, where here the final
state
$p$ specifies both nucleon and higher mass baryon excitation, is assumed to be
dominated by virtual photon exchange. It is therefore written in terms of two
structure functions $F_2^{D(4)}$ and $\frac{F_2^{D(4)}}{2x(1+R^{D(4)})}$ in the
form
($\alpha$ is here the fine structure constant)
\begin{equation}
\frac{{\rm d}^4 \sigma_{ep\rightarrow epX}}{{\rm d}x{\rm d}Q^2{\rm
d}x_{I\!\!P}dt}
= \frac{4\pi \alpha^2}{xQ^4}\,
  \left\{1-y+\frac{y^2}{2[1+R^{D(4)}(x,Q^2,x_{I\!\!P},t)]}\right\}
  \,F_2^{D(4)}(x,Q^2,x_{I\!\!P},t)
           \label{eq:defF2D}
\end{equation}
in which $y$ is the usual DIS scaling variable given by $y=Q^2/sx$. $s$ is the
$ep$ collision centre of mass (CM) energy squared.
It is convenient to express this cross section in terms of $F_2^{D(4)}$ and
$R^{D(4)}$ because the data available are
predominantly at low $y$ so that there is little sensitivity to $R^{D(4)}$. At
fixed $x_{I\!\!P}$ and $t$ for the range of $y$ of the results presented here
($y\leq 0.428$), $F_2^{D(4)}$ increases by no more than $17\%$ for
$0<R^{D(4)}<\infty$.

The measurement presented here uses data in which the final state colourless
remnant is not detected. Therefore no accurate determination of $t$ is possible
and the measured cross section amounts to
$\frac{d^3 \sigma (ep\rightarrow epX)}{{\rm d}x{\rm d}Q^2{\rm d}x_{I\!\!P}}$,
from which
it is possible to determine only
$F_2^{D(3)}(x,Q^2,x_{I\!\!P})=\int F_2^{D(4)}(x,Q^2,x_{I\!\!P},t)\,{\rm d}t$
provided that a particular choice is made for $R^{D(4)}$ and its $t$
dependence.
The integration is over the range
$\mid\!t_{min}\!\mid<\mid\!t\!\mid<\mid\!t\!\mid_{lim}$ where $t_{min}$ is a
function of $Q^2$, $W^2$, $M_X^2$ and the mass of the colourless
remnant, and $\mid\!t\!\mid_{lim}$ is specified by the
requirement that all particles in the colourless remnant remain undetected
(see section~\ref{sec-datastuff}). For this measurement, $R^{D(4)}$ is set
to $0$ for all $t$ and $F_2^{D(3)}$ is evaluated from
\begin{equation}
\frac{{\rm d}^3 \sigma_{ep\rightarrow epX}}{{\rm d}x{\rm d}Q^2{\rm
d}x_{I\!\!P}}
= \frac{4\pi \alpha^2}{xQ^4}\,
  \left\{1-y+\frac{y^2}{2}\right\}
  \,F_2^{D(3)}(x,Q^2,x_{I\!\!P})
           \label{eq:F2D3}
\end{equation}
following the original procedure of~\cite{Ingprytz1,Ingprytz2}.

The above definition of $F_2^{D(3)}(x,Q^2,x_{I\!\!P})$ renders
quantitative comparison with the structure function
$F_2(x,Q^2)$ of the proton straightforward
{}~\cite{H1F293}.

\section{H1 Apparatus and Kinematic Reconstruction}

The data were taken with the H1
detector at the HERA $ep$ collider at DESY
in 1993, in which $26.7\,{\rm GeV}$ electrons were
in head--on collision with $820\,{\rm GeV}$
protons, $\sqrt{s}= 296\,{\rm GeV}$.
The H1 detector is described in more detail in~\cite{H1rapgap,H1det,LAC1,Fmu}.
Here the most important aspects
necessary to explain the procedures for the analysis are summarised.

In the following, a coordinate system is
used with origin at the interaction point and $z$ axis along
the proton beam, or forward, direction. The pseudo--rapidity of a final state
particle with polar angle $\theta$ in the laboratory is then
$\eta = -\ln \tan \frac{\theta}{2}$.

Scattered electrons are measured in the backward electromagnetic calorimeter
(BEMC, electromagnetic energy resolution
$\sigma_{E}/E \sim 10\%/\sqrt{E\,{\rm (GeV)}}$ and overall
scale known to within $1.7\%$) and in a multi--wire proportional chamber (BPC).

Charged particles are detected in
central and forward tracking detectors (CTD and FTD). They consist of drift
chambers interspersed with multi--wire proportional chambers for fast signals
for
triggers using charged particles originating from the $z$ range of the event
vertex.

Hadronic energy flow in the event final state is measured in the liquid
argon (LAr) calorimeter ($-1.51<\eta<3.65$). The hadronic energy resolution is
$\sigma_E/E \sim 50\%/\sqrt{E\,{\rm (GeV)}}\,\oplus 2\%$ (measured
in a test beam~\cite{LAC2}) and the overall scale is known to within
$6\%$.

The selection of events in which there is no energy flow adjacent to the
direction of any proton remnant requires detectors
with the best possible coverage in the forward region.
The forward ``plug" calorimeter (PLUG; $3.54<\eta<5.08$) is
used in this analysis to ``tag" the production of energy above a threshold of
$1\,{\rm GeV}$. The
forward muon detector (FMD) is used in this analysis to ``tag" the
production of hadrons in the forward pseudo--rapidity range $5.0<\eta<6.6$ by
detecting and reconstructing track segments due to charged particles produced
in
secondary interactions of these hadrons in the collimators, the beam pipe,
and adjunct material~\cite{Mehta,Phillips}.

Two electromagnetic calorimeters (LUMI) situated downstream in the electron
beam
direction measure
electrons and photons from the process $ep\rightarrow ep\gamma$ for the purpose
of luminosity determination. They are also used in this analysis to help
quantify photoproduction background.

The kinematic quantities $x$ and $Q^2$ were reconstructed using the ``$\Sigma$
method"~\cite{Bernardi,H1F293}, which ensures acceptable resolution across
the entire kinematic range considered. The procedure, which is relatively
insensitive to initial state radiative effects, uses information both from the
scattered electron reconstructed using the BEMC and the BPC and from the
hadronic energy
deposition in the LAr calorimeter and BEMC, together with the event vertex. No
attempt
was made to use energy flow detected in PLUG and FMD for the purposes of
reconstructing the kinematic variables of events.

The mass of the hadronic system excluding the colourless remnant, $M_X$, was
determined directly from the calorimeter cells in the LAr calorimeter and BEMC
and from the event vertex determined using the tracks reconstructed in the CTD
and in the FTD. $\beta$ was then determined using $Q^2$ from above and $M_X$
substituted in equation~(\ref{eq:beta}). $x_{I\!\!P}$ then follows using
equation~(\ref{eq:trivial}).

\section{Data Taking and Event Selection}
\label{sec-datastuff}

The data were obtained with a trigger requiring the
presence of a localized energy cluster in the BEMC of $4\,{\rm GeV}$ or
more.
The selection procedure described in~\cite{H1F293} was used to obtain a sample
of DIS $ep$ events, which were then constrained to the kinematic ranges
$7.5<Q^2<70\,{\rm GeV^2}$ and $0.03<y<0.7$.
A total of $16366$ events satisfied the selection criteria from an integrated
$ep$ luminosity of $271\pm 14\,{\rm nb}^{-1}$.
\begin{figure}[tb]  \centering
\vspace*{-1.4cm}
\begin{picture}(0,0)(0,0)
\put(-57,180){a)}
\put(19,180){b)}
\put(-57,135){c)}
\put(19,135){d)}
\put(-57,90){e)}
\put(19,90){f)}
\put(-57,45){g)}
\put(19,45){h)}
\end{picture}
\vspace*{-1.0cm}
\caption{\footnotesize Distributions of a) $\eta_{max}$ measured in the LAr
calorimeter for the sample of DIS events; b) the number of
charged track segments reconstructed in FMD $N_{{\rm FMD}}$ per event for DIS
events with $\eta_{max}>3.2$; c) total energy deposition per
event in the PLUG for DIS events with $\eta_{max}>3.2$; d) hadronic
invariant mass of the final state system $M_X$ for the selected rapidity gap
(RG) events; e) $\eta_{max}$
for DIS events and for the rapidity gap events with the ``standard DIS"
background
subtracted; f) $y$, g) $x_{I\!\!P}$, and h) $\beta$ for the
rapidity gap events with the DIS background
subtracted; in a), b), c) and d) the expectations of
``standard DIS" simulations (CDM, MEPS), normalised to the data for
$\eta_{max}>3.2$, are also shown and the errors are
statistical; in e), f), g) and h) the
expectations of the Monte Carlo simulation (DIFF) used to
extract the contribution of rapidity gap events in the form of $F_2^{D(3)}$
are also shown, and the systematic error due to due to the  DIS background
subtraction are included with the statistical error.}
\label{fig001}
\end{figure}

Figures~\ref{fig001}a), b) and c) show spectra which illustrate the response of
H1 detectors to these DIS events. Superimposed are two simulated
expectations, henceforth referred to as ``standard DIS", which are based on a
partonic picture and which describe the gross features of DIS data at
HERA~\cite{H1F293,Engyflow,Jets}. They
are LEPTO~\cite{Lepto} (MEPS), which utilises
$O(\alpha_S)$ matrix elements in QCD and parton showers, and a combination of
LEPTO and ARIADNE~\cite{Ariadne} (CDM) in which gluons are radiated from a
colour dipole. The parton distribution functions for the proton were taken from
MRS(H)~\cite{MRS(H)}. The $(x,Q^2)$ dependence of $F_2(x,Q^2)$ was adjusted to
reproduce the most recent measurement of the proton structure
function~\cite{H1F293}.

For the LAr calorimeter
(figure~\ref{fig001}a), the distribution of $\eta_{max}$, the pseudo--rapidity
of the most forward energy cluster~\cite{LAC1} with energy greater than
$0.4\,{\rm GeV}$, is shown. In all following
comparisons and analysis, the normalisations of the ``standard
DIS" simulations are fixed to the number of observed events with
$\eta_{max}>3.2$. Comparison with the ``standard DIS" expectations not
only demonstrates the excess of events with an interval of
pseudo--rapidity adjacent to the proton remnant which is devoid of hadronic
energy ($\eta_{max}\,\lapprox\,1.8$), as already reported in~\cite{H1rapgap},
but also illustrates how well the LAr calorimeter response and cluster analysis
are understood in the forward region ($\eta_{max}\,\gapprox\,1.8$).
In figure~\ref{fig001}b) the response of the FMD detector, taken to be the
number of track segments reconstructed in a pair of FMD drift chamber layers
($N_{FMD}$), for events with $\eta_{max} > 3.2$, is compared with
``standard DIS" expectation. For the PLUG calorimeter (figure~\ref{fig001}c),
the
energy spectrum ($E_{PLUG}$) for these same events is shown, together
with ``standard DIS" expectation. It can be seen that the measured responses in
H1 of FMD
($N_{FMD}$) and of PLUG ($E_{PLUG}$), as well as
of the LAr calorimeter ($\eta_{max}$) in the forward region
($\eta\,\gapprox\,3$), are all very well reproduced by the
``standard DIS" simulations for those DIS events with no large rapidity gap.

A sample of events, henceforth referred to as ``rapidity gap" events, in which
no final state hadronic energy flow is detected
adjacent to the proton remnant direction and which is not
reproduced by the ``standard DIS" simulations, was selected by requiring
$E_{PLUG}< 1\,{\rm GeV}$, $N_{FMD}\leq 1$, and
$\eta_{max}<3.2$. These threshold values for
$E_{PLUG}$ and $N_{FMD}$ were determined using randomly triggered events in H1,
for which
the responses of each are then due predominantly to noise coincidences and out
of time track segments respectively. In the rapidity gap sample, the remaining
background
of events which can be described by ``standard DIS" simulation was taken to be
that due to the CDM simulation because recent measurements of forward energy
flow in DIS are known to be well reproduced by it~\cite{Engyflow}.
The uncertainty in this background was estimated using the MEPS simulation,
which does not describe well the forward energy flow in DIS. Crucial to this
background subtraction are the efficiencies for rejection of events with
forward energy
flow. They depend on a detailed understanding of the performances of the LAr
calorimeter, PLUG, and FMD as ``taggers" of forward energy flow. A systematic
study of the response of each detector as a function of the presence or absence
of signals in the other two and comparison with simulation (CDM and MEPS)
yielded a self--consistent set of efficiencies with a quantifiable
error~\cite{Mehta}.

In figure~\ref{fig001}d) the distributions in $M_X$ are shown, for the rapidity
gap sample, and
for the ``standard DIS" background (CDM and MEPS). Using Monte Carlo simulation
of diffractive DIS
events (RAPGAP)~\cite{Rapgap}, the resolution $\sigma (M_X)/M_X$ in
$M_X$, determined using LAr and BEMC calorimeter cells, was found to be
$\leq 30\%$ for $M_X\geq 6\,{\rm GeV}$ and to rise to $\sim 100\%$ as $M_X$
decreased from $6\,{\rm GeV}$ to threshold (twice the pion mass). The
``standard DIS" background remaining after the rapidity gap selection can be
seen to vary between $10\%$ and about $50\%$ for $M_X$ less than about
$25\,{\rm GeV}$.
In the following quantitative analysis it is subtracted
bin by bin, and a contribution to the systematic uncertainty in the yield
of rapidity gap events is taken equal to the difference between the background
subtraction calculated using MEPS and that calculated using CDM.

Background, amounting to less than $1\%$ in total, in the rapidity gap sample
due to beam gas and beam wall interactions, LAr pile up, cosmic rays, two
photon
processes and Compton scattering were considered and removed or corrected for
as
stated in~\cite{H1rapgap}. The level of background due to photoproduction
interactions in the rapidity gap sample was determined to be $(5\pm 5)\%$
in the bin of lowest $x$ and $Q^2$, to be less in other bins, and to be in
total $1\%$, both using Monte Carlo simulation based on the
PYTHIA~\cite{PYTHIA}
and POMPYT~\cite{POMPYT} codes, and using those events in the rapidity gap
sample in which an electron was detected in the LUMI detectors.

The rapidity gap sample thus selected and used in the analysis in
section~\ref{sec-Measurement} below, amounted to
$1723$ events, and $1451$ after subtraction of all backgrounds. In
figure~\ref{fig001}e) the distribution of $\eta_{max}$ is shown both for all
DIS
events (as in figure~\ref{fig001}a) and
for the event sample after subtraction of the ``standard DIS" background (used
to determine $F_2^{D(3)}$ - see below). The errors in the distribution of
$\eta_{max}$ in figure~\ref{fig001}e) include the
systematic uncertainties for the selection based on the forward
detectors FMD, PLUG and LAr calorimeter ($\eta_{max}<3.2$) with the
``standard DIS" subtraction.

Below $\eta_{max}$ of $1.8$ the distribution of the rapidity gap events
in $\eta_{max}$ is everywhere below that for the total DIS sample. This
shortfall amounts to $21\%$ in total, of which $(7\pm 3)\%$ is due to
PLUG noise and out of time track segments which fake either or both of
energy deposition in the PLUG and more than one reconstructed segment in FMD,
and $(5\pm 3)\%$ is due to ``standard DIS" background. The remaining $(9\pm
4)\%$ is
attributed to proton dissociation, secondary particles from which
sometimes generate hits in FMD and PLUG. Monte Carlo simulation of DIS
diffraction, in which the incident proton dissociates, confirms this
interpretation at the observed rate
if roughly one third of simulated deep--inelastic diffractive events are
assigned
as proton dissociation.

The use of a forward rapidity gap to select events in which the proton
remnant is colourless also specifies the kinematic range in $t$ of the
measurement presented here. In the selected sample of events no particle from
the colourless remnant, be it nucleon alone or one of the dissociation products
of
the proton, may be detected in the detector with the
most forward sensitivity, namely FMD ($5.0<\eta<6.6$). It follows that
$\mid\!t\!\mid$ is less than roughly $7\,{\rm GeV}^2$ for the range
$x<x_{I\!\!P}<\xpomup$ of this measurement.

\section{Measurement of $F_2^{D(3)}(\beta,Q^2,x_{I\!\!P})$}
\label{sec-Measurement}

The rapidity gap event sample has been used to measure
$F_2^{D(3)}(\beta,Q^2,x_{I\!\!P})$ from the measured cross section
by means of equation~(\ref{eq:F2D3}). The results
were obtained from the data in the form of $F_2^{D(3)}$ as a function of
$\beta$, $Q^2$ and $x$, from which the dependence on $\beta$, $Q^2$ and
$x_{I\!\!P}$ follows using equation~(\ref{eq:trivial}).

Monte Carlo simulations based on deep inelastic
electron--pomeron ($eI\!\!P$) scattering (RAPGAP)~\cite{Rapgap} and on
diffractive elastic vector meson electroproduction~\cite{List,H1rapgap}
(based on~\cite{VMD,Yennie}) were used for the determination of acceptance and
experimental bias. Both have been demonstrated as necessary to describe well
the
main features of the data sample~\cite{H1rapgap}. Because the resulting sample
is selected on the basis of laboratory pseudo--rapidity $\eta$ in the
forward direction, it is confined to small values of $x_{I\!\!P}$
by virtue of a strong correlation between $\eta_{max}$ and
$x_{I\!\!P}$~\cite{Phillips}. Therefore $F_2^{D(3)}$ is evaluated for the well
defined range of the kinematic variable $x_{I\!\!P}<\xpomup$.

$F_2^{D(3)}$ was determined using the measured
event numbers in three dimensional bins of $\beta$, $Q^2$ and $x$. The bin
acceptances were calculated using Monte Carlo simulation to obtain bin averaged
cross sections~\cite{Mehta,Phillips}. The Monte Carlo simulation modelled
deep--inelastic $eI\!\!P$ scattering, in
which $I\!\!P$ structure was taken to be due to a ``hard" quark distribution
$x_{q/I\!\!P}(1-x_{q/I\!\!P})$ with admixtures of a softer gluon distribution
$(1-x_{g/I\!\!P})^5$~\cite{Rapgap,Ingprytz1,Ingprytz2}. The peripheral $t$
dependence for the $I\!\!P$ flux ``in" the proton followed the parameterisation
of~\cite{Streng}. Elastic vector
meson production was included following~\cite{List}. Here $x_{i/I\!\!P}$,
$i=q,g$ are the fractions of
the $4$--momentum of the $I\!\!P$ carried by the parton $i$ in the
$I\!\!P$'s structure. When the extracted bin averaged cross
sections were included in the Monte Carlo simulation above, the distributions
of
observables in H1 were described well. Examples are shown as the curves (DIFF)
in figures~\ref{fig001}e), f), g) and h), where the $eI\!\!P$ interaction mix
is
``hard quark" : ``soft gluon" : ``elastic VMD" $72:21:7\%$ in observed events.
Note
that the quoted proportion
of soft gluon events is for the choice of sub--process kinematic cut--off
$\hat{p}_T>0.3\,{\rm GeV}$.

The criteria used to choose the kinematic
boundaries of each bin included the minimisation of effects due to experimental
resolution and smearing for $\beta$ and $x$, and adequate statistics for $Q^2$.
No bin with acceptance less than $30\%$ was included. The application of a
selection cut $\eta_{max}<3.2$ for the sample of rapidity gap events ensured
that
the invariant mass $M_X$, and therefore $\beta$, was well reconstructed. The
diffractive Monte Carlo was used to choose bins in $\beta$ for which the
average error in $\beta$
never exceeded $65\%$ of the width of the bin, and for which the systematic
shifts from true $\beta$ were
always $< 5\%$~\cite{Mehta,Phillips}. None of the bins in $\beta$ included the
mass range corresponding to the electroproduction of the vector mesons
$\rho(770)$, $\omega(783)$ and $\phi(1020)$. Each bin averaged cross section
was then
interpolated to a bin centre value in $(\beta,Q^2,x)$~\cite{Phillips}, and
$F_2^{D(3)}$ was then calculated by straightforward application of
equation~(\ref{eq:F2D3}).

Details concerned with the evaluation of all sources of systematic error in the
extraction of $F_2^{D(3)}$ are in~\cite{Mehta,Phillips}. They are quoted below
as average contributions to $F_2^{D(3)}$ so as to give an indication of their
relative significance:
\begin{itemize}
\item radiative corrections: $3\%$;
\item uncertainty in acceptance calculation due to absence of higher order QED
terms in Monte Carlo simulation: $6\%$;
\item ``standard DIS" background subtraction: $17\%$;
\item uncertainties in the diffractive simulation for acceptance correction due
to a priori ignorance of physics sub--processes: $11\%$
\item $6\%$ uncertainty in the LAr calorimeter hadronic energy scale: $6\%$;
\item $1.7\%$ uncertainty in the BEMC electromagnetic energy scale: $4\%$;
\item $2\,{\rm mrad}$ uncertainty in electron scattering angle $\theta_e$:
$4\%$
\item uncertainty in experimental acceptance due to a priori ignorance of
$F_2^{D(3)}$: $8\%$.
\end{itemize}
Having been evaluated for each bin, the above sources of error are combined to
form the systematic errors quoted in the results in table~\ref{table1}.
To include both proton scattering and proton dissociation,
an overall correction was made for the shortfall $(9\pm 4)\%$ attributed to
dissociation (see section~\ref{sec-datastuff}). Therefore
$F_2^{D(3)}$ is
evaluated as the total diffractive contribution to the proton structure
function $F_2$.

The results for $F_2^{D(3)}$ are shown as the $(\beta,Q^2,x_{I\!\!P})$
dependence in figure~\ref{fig002} and as the $(\beta,Q^2,x)$ dependence
in table~\ref{table1}. Note that the smallest total error on $F_2^{D(3)}$ is
$27\%$. Therefore there is no sensitivity to any choice of
$R^{D(4)}$ in equation~(\ref{eq:defF2D}) and no significant effect due to the
assumption $R^{D(4)}=0$ and the use of equation~(\ref{eq:F2D3}).
An overall uncertainty of $8\%$ is not included in the table. It
arises from the combination of three uncertainties, in luminosity measurement,
in the proportion of proton dissociation in rapidity gap events, and in the
selection of rapidity gap events because of PLUG calorimeter noise
and random FMD pairs. The results quoted use the ``$\Sigma$ method" for
kinematic reconstruction. The analysis has also been completed
using only the scattered electron to determine the kinematics and the results
agree to well within experimental
error. Repeating the whole analysis using different cuts, for example no
restriction on $\eta_{max}$, produced results for $F_2^{D(3)}$ which did not
differ to within systematic error from those quoted.
\begin{figure}[tb] \unitlength 1mm
\vspace*{-1.4cm}
\centering
\vspace*{-0.3cm}
\caption{\footnotesize
The diffractive contribution $F_2^{D(3)}(\beta,Q^2,x_{I\!\!P})$ to the
proton structure function $F_2$ as a function of $x_{I\!\!P}$ for different
$\beta$ and $Q^2$; the inner error bar is the statistical error; the full error
shows the statistical and systematic error added in quadrature; superimposed is
the result of the fit establishing a
factorisable dependence of the form $\propto x_{I\!\!P}^{-n}$ (see text).
Note that an overall normalisation uncertainty of $8\%$ is not included.}
\label{fig002}
\end{figure}

\vspace*{-0.3cm}

\section{Factorisation of
$F_2^{D(3)}$ and Evidence for Diffraction}
\label{sec-factorisation}

Everywhere $F_2^{D(3)}$ is observed to decrease
monotonically with increasing $x_{I\!\!P}$ in the measured range
$\xpomlo \leq x_{I\!\!P}<\xpomup$. An excellent fit to all data points,
irrespective of $\beta$ and $Q^2$, is obtained assuming a polynomial dependence
$x_{I\!\!P}^{-n}$
with a single exponent $n=\n$, $\chi ^2/{\rm d.f.}=32.0/46$, $94\%$ C.L.
($\chi ^2/{\rm d.f.}=64.5/46$, $4\%$ C.L. assuming only statistical errors).
The
observed universal dependence
on $x_{I\!\!P}$ is thus a feature of the measurements at the level of
statistical accuracy. There is no evidence for any systematic trend in the
contributions to $\chi ^2$ as a function of $\beta$ and $Q^2$. Using only the
reconstructed electron to determine $x$ and $Q^2$ an otherwise identical
analysis yielded $n=\nel$.

Such a universal dependence on $x_{I\!\!P}$, independent of $\beta$ and $Q^2$,
is expected naively if the deep--inelastic process responsible for rapidity gap
events involves a (colourless) target $T$ in the incident proton whose
characteristics are not dependent on $x_{I\!\!P}$, and which carries only a
small fraction of the proton's momentum. The dependence of
$F_2^{D(3)}(\beta,Q^2,x_{I\!\!P})$ then factorises into the product of a
universal term ($f_{T/p}(x_{I\!\!P})\propto x_{I\!\!P}^{-n}$), which
describes the ``flux" of $T$ ``in" the proton, and a term which describes any
structure of $T$ and which is a function of only $\beta$ and $Q^2$.
The electron scattering cross section for the process $ep\rightarrow eXp,N^*$
etc. can then be written
\begin{equation}
\frac{{\rm d}^3 \sigma _{ep\rightarrow epX}}{{\rm d}\beta {\rm
d}Q^2dx_{I\!\!P}}
=\frac{{\rm d}^2 \sigma _{eT\rightarrow eX}}{{\rm d}\beta {\rm d}Q^2}\cdot
  f_{T/p}(x_{I\!\!P})
           \label{eq:xfact}
\end{equation}
where $\frac{{\rm d}^2 \sigma _{eT\rightarrow eX}}{{\rm d}\beta {\rm d}Q^2}$
describes the deep--inelastic scattering of the electron off $T$.
\begin{table}[tb]
\vspace*{-1.4cm}
\begin{scriptsize}
\begin{picture}(100,150)(0,0)
\put(5,80){\begin{tabular}{|c|c|c|c|c|c|c|} \hline
\multicolumn{7}{|c|}{$Q^2=8.5$~${\rm {\rm GeV}}^2$} \\ \hline
 $\beta$ & $x$ & $F_2^{D(3)}$ & ST& SY & AC & FB\\ \hline
\multicolumn{7}{|c|}{$Q^2=12.0$~${\rm {\rm GeV}}^2$} \\ \hline
 $\beta$ & $x$ & $F_2^{D(3)}$ & ST& SY & AC & FB\\ \hline
\end{tabular}}
\put(85,78.2){\begin{tabular}{|c|c|c|c|c|c|c|} \hline
\multicolumn{7}{|c|}{$Q^2=25.0$~${\rm {\rm GeV}}^2$} \\ \hline
 $\beta$ & $x$ & $F_2^{D(3)}$ & ST& SY & AC & FB\\ \hline
\multicolumn{7}{|c|}{$Q^2=50.0$~${\rm {\rm GeV}}^2$} \\ \hline
 $\beta$ & $x$ & $F_2^{D(3)}$ & ST& SY & AC & FB\\ \hline
\end{tabular}}
\end{picture}
\end{scriptsize}
\vspace*{-1.7cm}
\caption{\footnotesize
The diffractive contribution $F_2^{D(3)}$ to
the proton structure function $F_2$ as a function of $\beta$, $Q^2$ and $x$; ST
is the statistical error,
SY is the total systematic error,
AC is the (smeared) acceptance, and FB the proportion of the signal
calculated to be background (see text). Not
included in the errors is an overall normalisation uncertainty of $8\%$.}
\label{table1}
\end{table}

If the colourless component $T$ of the proton is taken to be well
parameterised as a ``Reggeised" hadronic exchange, then the $x_{I\!\!P}$
dependence of the flux factor is specified by the leading Regge trajectory
$\alpha(t)$ in the ``asymptotic limit" $x_{I\!\!P}\rightarrow 0$,
$t/s \rightarrow 0$. The dependence takes the form
$x_{I\!\!P}^{-[2\alpha(t)-1]}$~\cite{Regge,Chew,Collins}.
In this measurement we perforce integrate over $t$. To good approximation,
$\alpha(t)$ may be replaced by $\alpha_{t=0}$ because the $t$ distribution of
the
soft hadronic proton interaction is likely to be ``peripheral" and the $t$
dependence of $\alpha(t)$ is likely to be slight.

The observed $x_{I\!\!P}$ dependence corresponds to
$\alpha_{t=0}=\ALPHA$. It is therefore consistent with the leading (effective)
trajectory which describes phenomenologically ``soft hadronic" diffractive
interactions, namely the $I\!\!P$ with $\alpha(t)=\alpha_{t=0}+\alpha' t$ and
$\alpha_{t=0}=1.085$, $\alpha'=0.25\,{\rm GeV}^{-2}$~\cite{ppdiff,Softpom}. If
the $t$ distribution of
proton diffraction is here like that in soft $pp$ collisions, namely peripheral
with a dependence $e^{bt}$ with $b > 1$, then the systematic
error, which arises by ignoring any $t$ dependence of $\alpha(t)$ of size given
by the slope $\alpha'=0.25\,{\rm GeV}^{-2}$ of the $I\!\!P$ trajectory, is
$<4\%$.
Note that the leading meson Regge trajectories, the $f_2(1270)/\omega(783)$ for
a
diffractive--like process with no isospin exchange, and the exchange degenerate
$\rho(770)/a_2(1320)$ for a process involving isospin exchange, have intercept
$\alpha_{t=0}\sim 0.5$, giving rise to an $x_{I\!\!P}$ dependence which is much
less steep. Parameterisations of pion exchange, either ``Reggeised"
($\alpha_{t=0}\sim 0$) or otherwise, have only a slight dependence on
$x_{I\!\!P}$.  Therefore the interpretation of the rapidity gap
events in DIS at low Bjorken--$x$ as being due in the main either to
diffractive
scattering or to diffractive dissociation of the incident proton can be made
without ambiguity.

The $x_{I\!\!P}$ dependence observed here does not specify the $M_X^2$
dependence in the process $ep\rightarrow epX$ without a further assumption for
the $\beta$ dependence of
$\frac{{\rm d}^3 \sigma _{ep\rightarrow epX}}{{\rm d}\beta {\rm d}Q^2 {\rm
d}x_{I\!\!P}}$.
However, note that $x_{I\!\!P}=M_{eX}^2/s$ where $M_{eX}^2$ is the invariant
mass squared recoiling against the proton in the inclusive process
$pe\rightarrow p(eX)$. Therefore the observation of an $x_{I\!\!P}$ dependence
$\propto x_{I\!\!P}^{-n}$ in the process $ep\rightarrow epX$ is
equivalent to a dependence on $M_{eX}^2$ of $(M_{eX}^2)^{-n}$ at fixed $ep$ CM
energy $\sqrt {s}$ in the inclusive process $pe\rightarrow p(eX)$. It is the
distribution in missing invariant mass squared $M_Y^2$ of the form
$\propto (M_Y^2)^{-[2\alpha(t)-1]}$, $\alpha_{t=0}=1.085$,
$\alpha'=0.25\,{\rm GeV}^{-2}$ in the soft hadronic interaction
$pp\rightarrow pY$ which is characteristic of high energy inelastic diffraction
and which specifies the effective $I\!\!P$ trajectory $\alpha(t)$ in $pp$
diffractive dissociation~\cite{Softpom}.

In calculations which use the BFKL QCD formalism of $I\!\!P$ dynamics, the
$I\!\!P$ is
found to be ``hard" with an intercept ($\alpha_{t=0} = 1+\varepsilon$,
$\varepsilon\,\lapprox\,0.5$)~\cite{Hardpom,BFKL}. Combining in quadrature the
statistical and systematic errors of the measurement of the $x_{I\!\!P}$
dependence above, the intercept cannot exceed $1.25$ (i.e. $\varepsilon \leq
0.25)$
with $99.7\%$ confidence. This does not rule out such a ``hard" $I\!\!P$ if
it contributes together with a soft component, such as that described above.
A recent calculation, based on a BFKL approach, suggests also that the simplest
factorisation with an $x_{I\!\!P}$ dependence $\propto x_{I\!\!P}^{-n}$ may not
strictly hold~\cite{NZ}. The accuracy of this measurement of the $x_{I\!\!P}$
dependence does not exclude such a possibility.

\section{The Deep--Inelastic Structure of the Pomeron}
\label{sec-DIdiff}

The observation that rapidity gap events in DIS are dominantly of a diffractive
nature, and the evidence for a simple factorisation of the
structure function $F_2^{D(3)}(\beta,Q^2,x_{I\!\!P})$ over values of
$x_{I\!\!P}$ in
the range $\xpomlo<x_{I\!\!P}<\xpomup$ for different values of $\beta$
and $Q^2$, lead naturally to the interpretation of the $\beta$ and $Q^2$
dependence of $F_2^{D(3)}$ as a measure of the deep--inelastic structure of
diffractive exchange. If such exchange is taken phenomenologically to be
described by an effective Regge trajectory, this interpretation amounts to the
deep--inelastic structure of the $I\!\!P$.

In figure~\ref{fig003} are shown the $\beta$ and $Q^2$ dependences of the
integral
\begin{equation}
\tilde{F}_2^D(\beta,Q^2)=
\int_{x_{I\!\!P_L}}^{x_{I\!\!P_H}} F_2^{D(3)}(\beta,Q^2,x_{I\!\!P})\,{\rm
d}x_{I\!\!P},
           \label{eq:F2Dint}
\end{equation}
which, assuming factorisation, is proportional to the structure function of the
$I\!\!P$. The range of integration, $x_{I\!\!P_L}=\xpomlo$ and
$x_{I\!\!P_H}=\xpomup$ is chosen to span the entire $x_{I\!\!P}$ range of
measurements of $F_2^{D(3)}$. $\tilde{F}_2^D$ is
evaluated assuming that the factorisation hypothesis with the
measured dependence on $x_{I\!\!P}$ is a good
description over this range of integration. The assumption is therefore made
that, for certain values of $\beta$ and $Q^2$,
the observed $x_{I\!\!P}$ dependence is valid beyond the measured range in
$x_{I\!\!P}$, including values which are kinematically inaccessible in this
measurement. Note therefore that, unlike $F_2^{D(3)}$ or any
integration of it over kinematically accessible values of
$\beta$, $Q^2$ and $x_{I\!\!P}$, $\tilde{F}_2^D$ is not a measure of
the diffractive contribution to $F_2$. The errors in the results for
$\tilde{F}_2^D(\beta,Q^2)$ include the contributions from the uncertainty above
(both statistical and systematic) in the ``flux" dependence on $x_{I\!\!P}$.
It is convenient to define $\tilde{F}_2^D(\beta,Q^2)$ in this way
to avoid the need to specify the normalisation of the unknown diffractive, or
$I\!\!P$, ``flux" in the proton when presenting the data in a form which is
suitable for direct comparison with theoretical expectation for $I\!\!P$
structure.
\begin{figure}[tb]
\vspace{-1.4cm}
\centering
\vspace{-1cm}
\caption{\footnotesize
Dependence of $\tilde{F}_2^D(\beta,Q^2)$ on $Q^2$ and $\beta$;
superimposed on the $Q^2$ dependence are the results of fits
at each $\beta$ which assume leading logarithmic scaling violations;
the best fit (continuous curve) and the curves corresponding to a change of
$\pm 1$ standard deviation (dashed curves) in the slope are shown; superimposed
on the $\beta$
dependence are the simplest $q{\bar q}$
expectation for $I\!\!P$ structure - [$\beta(1-\beta)$] (continuous curve),
and a constant dependence (dashed curve), for which the overall normalisations
are determined from fits to the data; also displayed is
a dependence of the form [$(1-\beta)^5$] (dotted curve) with arbitrary
normalisation. Note that an overall normalisation uncertainty of $8\%$ is not
included.}
\label{fig003}
\end{figure}

There is no evidence for any substantial $Q^2$ dependence of
$\tilde{F}_2^D$. Deep (high $Q^2$) diffractive interactions are thus broadly of
a scale invariant, and therefore point-like, nature. Furthermore,
$\tilde{F}_2^D$ is observed to have little dependence on
$\beta$ with contributions throughout the measured range ($0.065<\beta <
0.65$).
Therefore the structure which is resolved by the
electron in the high $Q^2$ diffractive interaction carries only a fraction of
the $I\!\!P$'s momentum, implying that the $I\!\!P$ has its own
sub--structure which may presumably be attributed to point-like partons.

A fit to the hypothesis of a linear $\log_{10} Q^2$ dependence demonstrates
that
the data are consistent with zero slope to within $\sim 1$ standard deviation
($\sigma$). The errors on the slopes in the fits shown in figure~\ref{fig003}
amount to
$\sigma=0.09\,(\log_{10}\,{\rm GeV}^2)^{-1}$ for all $\beta$ except that at
$\beta=0.065$ for which $\sigma=0.15\,(\log_{10}\,{\rm GeV}^2)^{-1}$. Therefore
the accuracy of this first measurement of the $Q^2$ dependence of
$\tilde{F}_2^D$ is such that the possibility
of ``scaling violations", symptomatic of QCD, may be admitted.
Furthermore, either they may be $\beta$ dependent and thus characteristic of
that observed for the structure function of a hadron such as the
proton, or they may be insensitive to $\beta$ and thus characteristic of the
structure function of a ``dressed" point--like field quantum such as the
photon.

Also shown in figure~\ref{fig003} superimposed on the $\beta$ dependence of
$\tilde{F}_2^D$ is the expectation based on the simplest $q{\bar q}$
picture of deep--inelastic diffraction, namely a ``hard" quark $\beta(1-\beta)$
dependence~\cite{Ingprytz1,Ingprytz2}. Such a dependence has been justified in
an approach in which $q$, ${\bar q}$ or both ``scatter" diffractively off the
proton~\cite{Softpom,NZ}. In such a picture, diffractive quark scattering at
high energy amounts to the exchange of two or more gluons and has a leading
order energy dependence of Regge form~\cite{LandNacht}. There is a suggestion
that
the observed $\beta$ dependence may exceed this simple $q{\bar q}$ expectation
as $\beta\rightarrow 0$, which is naturally to be expected
in any quantum chromodynamic interpretation which includes gluons as well as
quarks. A fit assuming no dependence of $\tilde{F}_2^D$ on
$\beta$ is equally acceptable. The ansatz of a soft dependence $(1-\beta)^5$
is completely ruled out. The lack of measurements at large $\beta$, where
elastic
vector meson production contributes, means that no conclusion can be drawn
concerning the existence or otherwise of a
``super--hard" component in $I\!\!P$ structure, first suggested by Brandt et
al.~\cite{Super}.

The dependence of
$\tilde{F}_2^D$ on $Q^2$ and on $\beta$, the appropriate Bjorken--$x$
like variable, resembles well established measurements of the
structure functions $F_2$ of hadrons in that it is consistent with an
understanding based on asymptotically free partons. It however contrasts with
them in that the Bjorken--$x$ dependences of the
structure functions $F_2$ of hadrons are all
observed to decrease substantially with increasing Bjorken--$x$. This suggests
the simplest possible interpretation for the sub--structure of the $I\!\!P$,
namely that due to two ``valence--like" partons which share the
majority of the $4$--momentum of the $I\!\!P$, together with
modifications at low $\beta$ due to QCD evolution.

\section{Summary and Conclusions}

The contribution to inclusive deep--inelastic electron--proton scattering (DIS)
of events, in which a region of pseudo--rapidity adjacent to the proton remnant
direction is devoid of hadronic energy and which is not described in the
framework of our present partonic understanding of DIS, has been evaluated in
the
form of a ``diffractive structure function" $F_2^{D(3)}(\beta,Q^2,x_{I\!\!P})$.

The dependence of $F_2^{D(3)}$ on $x_{I\!\!P}$, which may
be interpreted as the fraction ($x_{I\!\!P/p}$) of the $4$--momentum carried by
the colourless component of the proton with which the
electron interacts, is measured to be $x_{I\!\!P}^{-n}$ with $n=\n$.
This dependence is found to be universal, irrespective of the deep--inelastic
scattering
variables $\beta$ and $Q^2$. This is as expected if the electron--proton
cross section can be factorised into the product of a term which describes
the ``flux" of colourless component of the proton and a term which corresponds
to the cross section for the interaction of the latter with the electron.

In the framework of ``Reggeised" hadronic exchange in the  $t$--channel which
couples to the incident proton, the dependence of
$F_2^{D(3)}$ on $x_{I\!\!P}$ may be interpreted as
the intercept $\alpha_{t=0}=\ALPHA$ of the leading Regge trajectory.
The latter is consistent with that of the pomeron trajectory
which describes phenomenologically the energy dependence of soft
hadronic diffraction, and is inconsistent with the intercepts
$\alpha_{t=0}\sim 0.5$ of the leading meson Regge trajectories. The origin of
``rapidity gap events" in deep--inelastic lepton nucleon scattering is
therefore demonstrated unambiguously to be predominantly diffractive, and the
colourless component of the proton to be consistent with the pomeron.
The  precision of the measurement of $\alpha_{t=0}$ does not exclude the
possibility of a contribution to deep--inelastic diffractive scattering from a
harder BFKL motivated pomeron trajectory with a higher intercept.

The deep--inelastic structure of the pomeron is observed to be consistent with
scale invariance for all measured values of the appropriate ``Bjorken--$x$"
like
variable $\beta$.
$F_2^{D(3)}$ is observed to be
non--zero for a significant range of values of $\beta$, demonstrating a
substantial inelastic contribution. The sub-structure
which is resolved in diffractive deep--inelastic
electron scattering thus carries only a fraction of the
momentum of the pomeron, meaning, when taken together with the
observation of broad consistency with
scale invariance, that this sub--structure may presumably be
attributed to point-like partons.

\section*{Acknowledgments}

We are grateful to the HERA machine group whose outstanding
efforts have made and continue to make this experiment possible. We thank
the engineers and technicians for their work in constructing and now
maintaining the H1 detector, our funding agencies for financial support, the
DESY technical staff for continual assistance, and the DESY directorate for the
hospitality which they extend to the non--DESY members of the collaboration.
We also acknowledge much help with the theoretical interpretation of our
results
from many colleagues, in particular G.\ Ingelman.

\end{document}